\documentclass{appolb}
\usepackage{hyperref}
\usepackage{graphicx} 
\usepackage{amsmath}
\usepackage{amssymb}
\usepackage{dsfont}

\newcommand{\tr}{\mathop{\mathrm{Tr}}}
\newcommand{\x}{\mathbf{x}}

\newcommand{\nom}{{\nonumber}}

\newcommand{\eref}[1]{Eq.~\eqref{#1}}
\newcommand{\secref}[1]{Sec.~\ref{#1}}

\newcommand{\figref}[1]{Fig.~\ref{#1}}

\begin{document}
\title{Phase diagram and isentropic curves from the vector meson
  extended Polyakov quark meson model
\thanks{Presented at Excited QCD 2017 (7\,-\,13 May 2017, Sintra, Portugal)}%
} \author{P{\'e}ter Kov{\'a}cs, Gy{\"o}rgy Wolf \address{Institute for
    Particle and Nuclear Physics, Wigner Research Centre for Physics,
    Hungarian Academy of Sciences, H-1525 Budapest, Hungary}
}

\maketitle
\begin{abstract}
  In the framework of the $N_f = 2+1$ flavor (axial)vector meson
  extended Polyakov quark meson model we investigate the QCD phase
  diagram at finite temperature and density. We use a $\chi^2$
  minimization procedure to parameterize the model based on
  tree\,-\,level decay widths and vacuum scalar and pseudoscalar
  curvature masses which incorporate the contribution of the
  constituent quarks. Using a hybrid approximation (mesons at tree
  level, fermions at one\,-\,loop level) for the grand potential we
  determine the phase boundary both on the $\mu_B-T$ and $\rho-T$
  planes. We also determine the location of the critical end point of
  the phase diagram. Moreover by calculating the pressure and other
  thermodynamical quantities derived from it, we determine a set of
  isentropic curves in the crossover region. We show that the curves
  behave very similarly as their counterparts obtained from the lattice
  in the crossover regime.
\end{abstract}

  
\section{Introduction}
\label{Sec:intro}

Our understanding of the properties of the strongly interacting matter
can be greatly improved with the upcoming FAIR facility in the sense
that investigation of a different region and might be the location of
the critical endpoint (CEP), if it exists at all, becomes
available. However, for the experiments better and better theoretical
predictions are also necessary.

With the help of a vector/ axial vector extended Polyakov quark meson
model, which was already described in \cite{elsm_T_2016}, we further
analyze some aspects of the chiral phase transition. In the present
approximation our model contains a scalar, a pseudoscalar, a vector
and an axial vector nonet field, beside the $u, d, s$ constituent
quarks. It is important to note the existence of a particle assignment
ambiguity in the scalar sector, namely there are more physical fields
below $2$~GeV in the scalar sector than we can describe with one
scalar nonet. Thus one have to try different assignments -- there are
$40$ possibilities -- during the determination of the Lagrangian
parameters and see which one is the best (for more detail see
\secref{Sec:param} and \cite{elsm_2013, elsm_T_2016}). With the best
set of parameters one can solve -- with some further approximations --
the field equations at finite temperature $T$ and/or baryon
chemical potential $\mu_B$. From that solution the phase boundary and
$T/\mu_B$ dependence of various physical quantities are readily given.

The paper is organized as follows. In \secref{Sec:model} the model is
introduced with its Lagrangian. Here the field equations, which
determine the temperature and baryochemical potential dependence of
the order parameters, are also presented. The parameterization is
briefly described in \secref{Sec:param}, followed by the results in
\secref{Sec:results} showing the phase boundary and a set of
isentropic curves compared with their lattice counterpart. Finally, we
conclude in \secref{Sec:concl}.

\section{The Model and the field equations}
\label{Sec:model}

The model is described by the following Lagrangian, which can be found
in more detail in \cite{elsm_2013, elsm_T_2016}:
\begin{align}
  \mathcal{L} & = \tr[(D_{\mu}M)^{\dagger}(D_{\mu}M)] -
  m_{0}^{2}\tr(M^{\dagger}M) - \lambda_{1}[\tr(M^{\dagger} M)]^{2}
  - \lambda_{2}\tr(M^{\dagger}M)^{2} \nom \\
  & + c_{1}(\det M+\det M^{\dagger}) + \tr[H(M+M^{\dagger})]
  - \frac{1}{4}\tr(L_{\mu\nu}^{2}+R_{\mu\nu}^{2}) \nom \\
  & + \tr\left[ \left(\frac{m_{1}^{2}}{2}+\Delta\right)
    (L_{\mu}^{2}+R_{\mu}^{2})\right] +
  i\frac{g_{2}}{2}(\tr\{L_{\mu\nu}[L^{\mu},L^{\nu}]\} +
  \tr\{R_{\mu\nu}[R^{\mu},R^{\nu}]\}) \nom \\
  & + \frac{h_{1}}{2}\tr(M^{\dagger}M)\tr(L_{\mu} ^{2} + R_{\mu}^{2})
  + h_{2}\tr[(L_{\mu}M)^{2}+(M R_{\mu} )^{2}] \\
  & + 2h_{3}\tr(L_{\mu}M R^{\mu}M^{\dagger}) + \bar{\Psi}\left[i
    \gamma_{\mu}D^{\mu}-g_F(\mathds{1}_{4\times 4} M_{S} + i\gamma_5
    M_{PS})\right]\Psi, \nom
\end{align}\label{Eq:Lagr}
with $M\equiv M_{S} + M_{PS}$ being the scalar\,--\,pseudoscalar
nonet fields, $L^{\mu}\equiv V^{\mu} + A^{\mu}, R^{\mu} \equiv V^{\mu} -
A^{\mu}$ stand for the left and right handed vector nonets (which are
linear combinations of the $V^{\mu}$ vector and $A^{\mu}$ axial vector
 nonet physical fields), the external fields are defined as $H =
\frac{1}{2}\textnormal{diag} (h_{0N},h_{0N}, \sqrt{2}h_{0S})$ and $\Delta
= \textnormal{diag} (\delta_{N}, \delta_{N}, \delta_{S})$, while
$G^{\mu} = g_s G^{\mu}_i T_i,$\footnote{Here $T_{i} = \lambda_{i}/2$
  ($i=1,\ldots,8$) denote the $SU(3)$ group generators, with the
  $\lambda_{i}$ Gell-Mann matrices} are the gluon fields.

As a usual process we use the spontaneous symmetry breaking scenario
with two non-zero vacuum expectation values denoted by $\phi_N$ and
$\phi_S$ -- connected to the $\lambda_N =
\frac{1}{\sqrt{3}}(\sqrt{2}\lambda_0+\lambda_8)$ non-strange and
$\lambda_S = \frac{1}{\sqrt{3}} (\lambda_0-\sqrt{2} \lambda_8)$
strange generators in the scalar sector. 

In the present -- mean field -- approximation we assume non-vanishing
gluon field only in the temporal direction ($G^4$), which is
additionally assumed to be x-independent and diagonal in color
space. This will give rise to the Polyakov loop variables $\Phi$ and
$\bar\Phi$, when calculating the partition function $\mathcal{Z}$ on
the constant gluon background $G^4$ (see details in
\cite{elsm_T_2016}). For the Polyakov-loop potential we use an
improved logarithmic potential (proposed in \cite{Haas_2013}), which
takes into account some part of the gluon dynamics.

The grand potential $\Omega$ is calculated by taking into account only
the fermionic fluctuations and treating the mesons at
three-level. Consequently, the grand potential has four parts, which
are the classical mesonic potential, the Polyakov-loop potential, the
fermionic vacuum and thermal parts,
\begin{equation}
  \Omega(T,\mu_q) =
  U_{\text{mes}}(\phi_N, \phi_S) + U_{\text{Pol}}(\Phi, \bar\Phi) +
  \Omega^{(0)\text{vac}}_{\bar q q} + \Omega^{(0)T}_{\bar q q}(T,\mu_q).
\end{equation}
Here $\mu_q=\mu_B/3$ and not every variable dependence is explicitly shown.

The values of the order parameters -- as a function of $T,\mu_q$ -- are
obtained from the four coupled field equations,
\begin{equation} 
\frac{\partial\Omega}{\partial \phi_N} =
\frac{\partial\Omega}{\partial \phi_S} =
\frac{\partial\Omega}{\partial \Phi} =
\frac{\partial\Omega}{\partial \bar\Phi} = 0.
\label{Eq:field}
\end{equation}
The detailed form of the field equations can be found in \cite{elsm_T_2016}. 

\section{Parameterization}
\label{Sec:param} 

To solve \eref{Eq:field} -- the set of coupled field equations -- one
has to determine the $14$ unknown parameters ($m_0$, $\lambda_1$,
$\lambda_2$, $c_1$, $m_1+\phi_N$, $h_1$, $h_2$, $h_3$, $\delta_S$,
$\phi_N$, $\phi_S$, $g_F$, $g_1$, $g_2$) of the \eref{Eq:Lagr}
Lagrangian. Therefore, we calculate a set of observables at zero
temperature (at tree-level: 5 vector/axial vector masses, 2 constituent
quark masses, 12 decay width, 2 PCAC relations, at one-loop level: 8
scalar/pseudoscalar curvature masses) and compare with their PDG
values \cite{PDG} through a multiparametric $\chi^2$ minimalization
procedure \cite{MINUIT}. Beside the zero temperature quantities we
also use the pseudocritical temperature $T_c$ at $\mu_B=0$ and compare
with its lattice result taken from \cite{aoki_2006}. It is worth to
note that the errors of the observables used for the $\chi^2$
minimalization are set to $20\%$ for the scalar masses and their decay
widths, $10\%$ for the constituent quarks and the $T_c$, and $5\%$ for
the remaining quantities as long as their PDG errors are not larger
(in that case we use the PDG error). We started the parameterization
procedure from $5\times 10^4$ points in the parameter space for the
different assignment scenarios of the scalar sector, and chose that
gave the minimal $\chi^2$ value. The resulting parameters and further
details can be found in \cite{elsm_T_2016}.


\section{Results: chiral phase boundary and isentropic curves}
\label{Sec:results}

\begin{figure}[htb]
  \centerline{\includegraphics[width=0.7\textwidth]{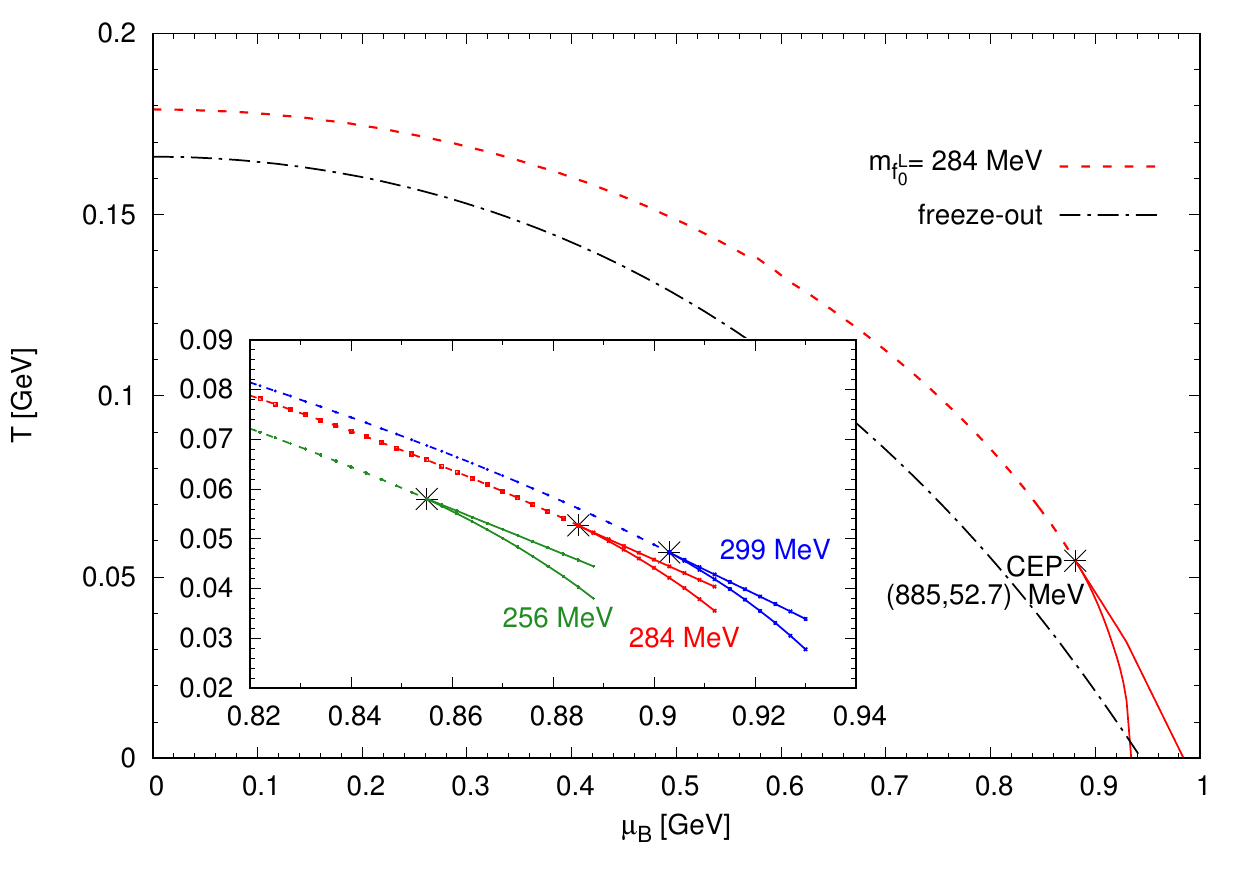}}
  \vspace*{-0.3cm}
  \caption{Phase diagram on the $T-\mu_B$ plane}
  \label{Fig:phase_diag_T_mu}
\end{figure}

Once \eref{Eq:field} is solved the points of the phase boundary are
given by the inflection points of the $\phi_N(T)$ curve for different
values of $\mu_B$. When solving the field equations we also calculate
the grand canonical potential $\Omega(T,\mu_q)$ along the solution
from which the pressure is simply given by
\begin{equation}
  p (T,\mu_q)=\Omega(T=0,\mu_q) - \Omega(T,\mu_q), 
\end{equation}
while the entropy and quark number densities are defined as
$s=\partial p/\partial T,$ and $\rho_q=\partial p/\partial \mu_q$
respectively. The phase diagram is shown in
\figref{Fig:phase_diag_T_mu} on the $T-\mu_B$ plane together with the
chemical freezout curve. On the phase boundary a second order critical
endpoint -- which separates the crossover and the first order regimes
-- exist at $(\mu_B,T) = (885,53)$~MeV. On the inset of
\figref{Fig:phase_diag_T_mu} the variation of the location of the CEP
with the $f_0$ mass is shown.
\begin{figure}[htb]
  \centerline{\includegraphics[width=0.7\textwidth]{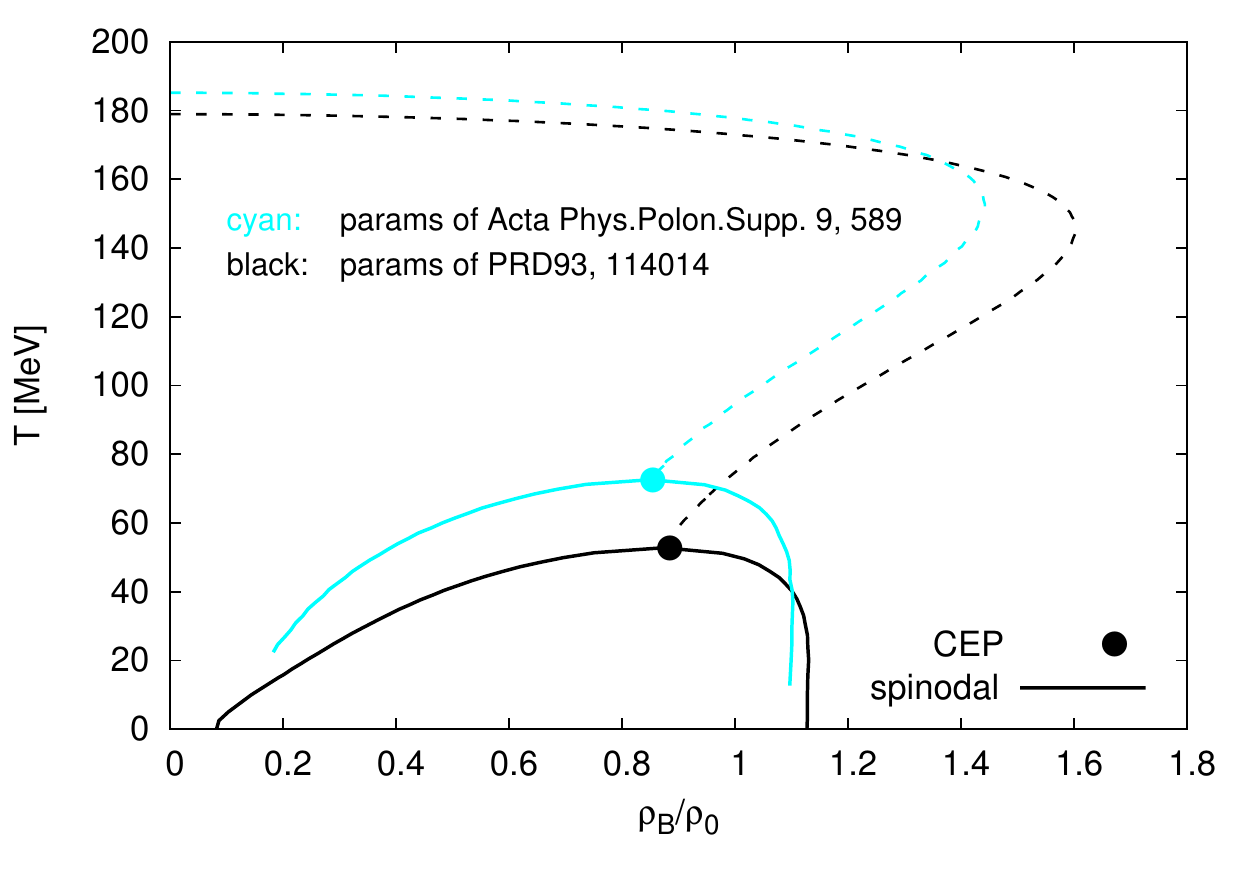}}
  \vspace*{-0.3cm}
  \caption{Phase diagram on the $T-\rho_B$ plane}
  \label{Fig:phase_diag_T_rho}
\end{figure} 
If instead of the baryochemical potential we use the $\rho_B=\rho_q/3$
baryon density we get the phase boundary on the $T-\rho_B$ plane
(\figref{Fig:phase_diag_T_rho}), where the baryon density is
normalized with the normal nuclear density $\rho_0 = 0.16
\frac{1}{\text{fm}^3}$. Here the two curves correspond to two
different Polyakov-loop potential presented in \cite{Szep:2016} (upper
curve) and \cite{elsm_T_2016} (lower curve). We find that the phase
boundary curves have a very similar shape found by others (see
e.g. \cite{Costa:2016}). However, our CEP is located at a lower
$\rho_B/\rho_0$ value than in \cite{Costa:2016}, where the CEP was
located at $\rho_B/\rho_0 \approx 2$ in a PNJL model calculation.

\begin{figure}[htb]
  \centerline{\includegraphics[width=0.7\textwidth]{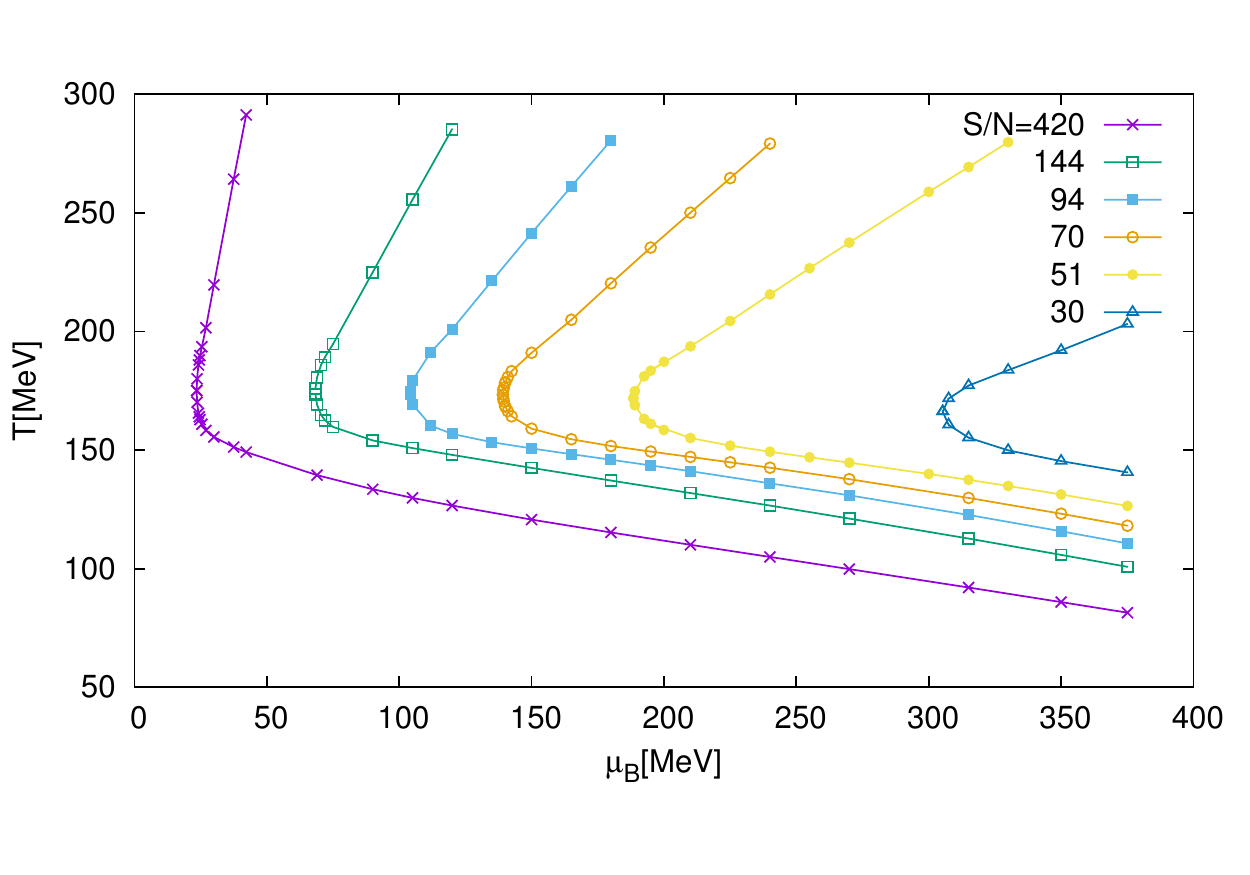}}
  \vspace*{-0.7cm}
  \caption{The calculated isentropic curves for different values of $S/N$.}
  \label{Fig:isentr}
\end{figure}

\begin{figure}[htb]
  \centerline{\includegraphics[width=0.7\textwidth]{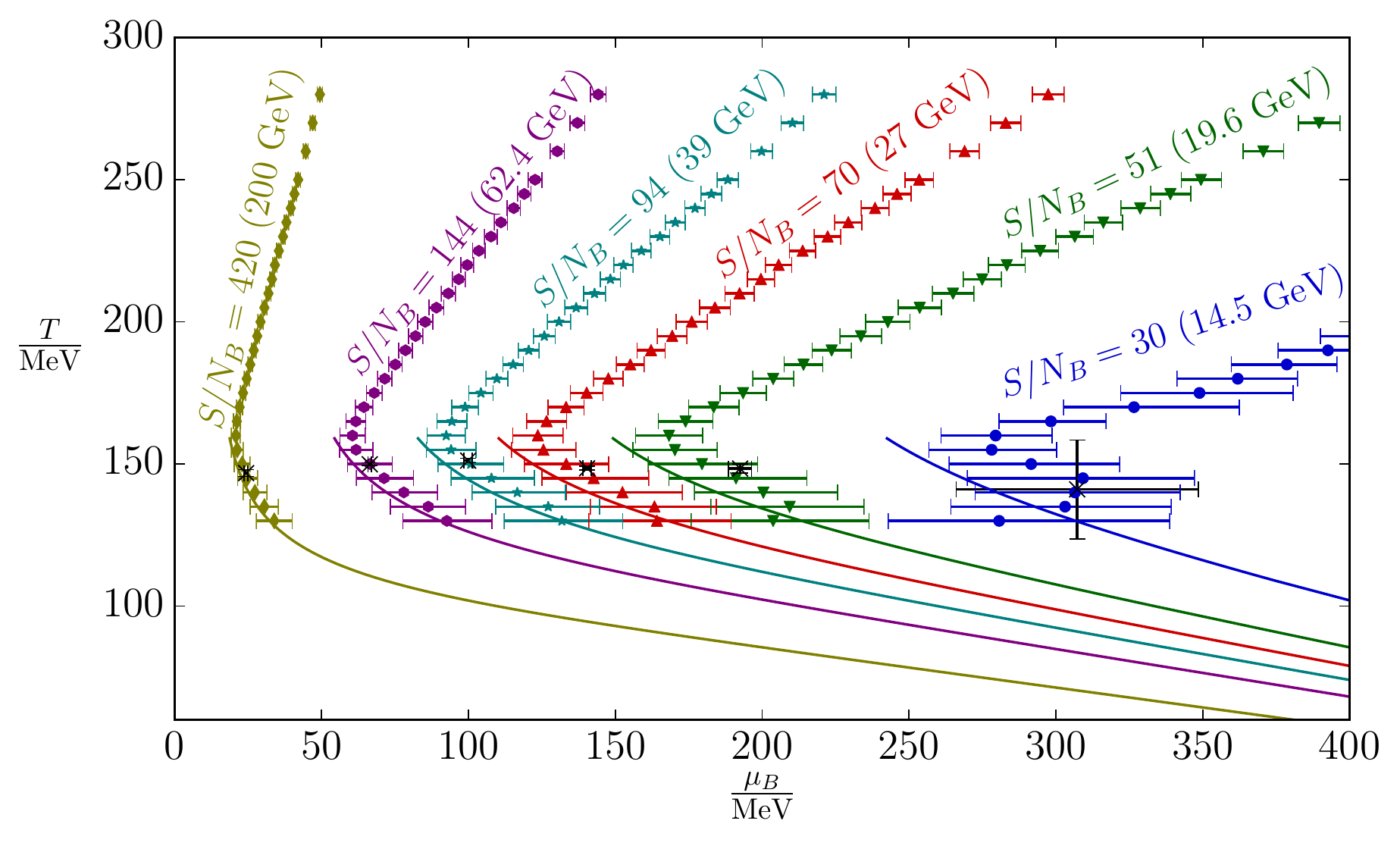}}
  \vspace*{-0.3cm}
  \caption{Isentropic curves from lattice for different values of
    $S/N$ from \cite{Gunther_2016}.}
  \label{Fig:lat_isentr}
\end{figure}

It is also interesting to investigate the isentropic curves on the
$T-\mu_B$ plane, which is defined as $S/N=s/\rho_q=\text{const.}$ These
curves can be determined on the lattice as well. Our curves are shown
in \figref{Fig:isentr}, while the lattice version, which uses analytic
continuation to the finite $\mu_B$ region is depicted in
\figref{Fig:lat_isentr} \cite{Gunther_2016}.
The two set of curves coincide very well. Since our curves are in the
crossover region and it is known -- from e.g. \cite{Costa:2016} --
that in the first order region the curves looks differently (always
have some 'S' shaped part), we can conclude that in the $\mu_B < 400$~MeV
region the existence of a CEP is unlikely on the lattice.

\section{Conclusion}
\label{Sec:concl}

In the framework of an (axial)vector Polyakov quark meson model some
aspects of the in medium properties of the strongly interacting matter
was presented. The calculated isentropic curves in the crossover regime
show a very good agreement with the corresponding curves on the
lattice. This suggest a large baryochemical potential value
($\mu_{B,\text{CEP}} > 400$~MeV) at the CEP on the lattice.

To improve our model, mesonic fluctuations and additional fields like
four-quark states could be included.

\section*{Acknowledgments}

The authors were supported by the Hungarian OTKA fund K109462 and by
the HIC for FAIR Guest Funds of the Goethe University Frankfurt. The
authors thank Z. Sz\'ep for making figures.

\end{document}